\documentclass[%
 reprint,
 amsmath,amssymb,
 aps,
]{revtex4-2}

\usepackage{graphicx}
\usepackage{dcolumn}
\usepackage{bm}

\begin{document}

\preprint{APS/123-QED}

\title{Optical chaos synchronisation in a cascaded injection experiment}

\author{Jules Mercadier$^{*,1,2}$}

\author{Yaya Doumbia$^{1,2}$}%
\author{Stefan Bittner$^{1,2}$}
\author{Marc Sciamanna$^{1,2}$}
\affiliation{%
 $^1$Chaire Photonique, LMOPS, CentraleSupélec, 57070 Metz, France\\
 $^2$Université de Lorraine, CentraleSupélec, LMOPS, 57070 Metz, France \\
 $^*$ Corresponding author : jules.mercadier@centralesupelec.fr
}%

\begin{abstract}
We experimentally study the synchronization of chaos generated by semiconductor lasers in a cascade injection configuration, i.e., a tunable master laser is used to generate chaos by optical injection in a transmitter laser that injects light into a receiver laser. Chaos synchronization between the transmitter and the receiver lasers is achieved with a correlation coefficient of 90 \% for a measurement bandwidth up to 35 GHz. Two parameter regions of good synchronization are found, corresponding to the alignment of the oscillation frequencies of the receiver laser with either the transmitter laser or the master laser.
\end{abstract}

\maketitle

\section{Introduction}
The synchronization of chaos generated by a laser has been studied for the so-called chaos cryptography \cite{sciamanna2015physics,VanWiggeren}  and the resulting secure communication has been demonstrated in fiber optic networks \cite{chen,argyris2005chaos}. Since the first observation of chaos synchronization by Pecora and Carroll in 1990 \cite{pecora1990}, various types of lasers have been used in synchronization experiments, including CO$_2$ lasers \cite{co2laser}, solid-state lasers \cite{solidstatelaser}, and semiconductor lasers (SLs) \cite{Slssynchro_1, 
 masoller_anticipativ}.  SLs can exhibit rich and diverse dynamics, including chaos, with the addition of an external disturbance, such as optical injection or feedback \cite{Wieczorek,simpson}. Most synchronization experiments conducted to date have employed chaos from a semiconductor laser with feedback. However, chaotic dynamics induced by delayed feedback requires sophisticated strategies for concealing the time delay signature in a chaos cryptography scheme \cite{rontani2007loss}. In addition, it is known that feedback-induced chaos has a bandwidth limited by the laser relaxation oscillation frequency, hence also limiting its use for broadband chaos synchronization.\\ 

In this paper, a cascade injection setup is explored, where optical injection from a first laser (master) generates broad-band chaos in the output of a second laser (transmitter). This signal is then injected into a third laser (receiver) to induce chaos synchronization. The properties of chaos synchronization in this setup have been less studied \cite{vaudel}. In \cite{vaudel}, this system was investigated with only 5 GHz measurement bandwidth, which may impact the correlation value as demonstrated below. Furthermore, only one chaotic regime was studied, with no information about the bandwidth of the generated chaos.

Here, we demonstrate excellent chaos synchronization with a correlation coefficient up to 90\% for a measurement bandwidth up to 35 GHz. Moreover, the cascade injection setup features two synchronization mechanisms between the transmitter and receiver laser: synchronization occurs when the receiver laser frequency either approaches the transmitter laser frequency or the master laser frequency. Finally, we compare the synchronization properties of the transmitter laser chaos when the chaotic dynamics is achieved by bifurcation from a wave-mixing dynamics or from undamped relaxation oscillations \cite{simpson,bistability}, and we observe good synchronization in both cases.
\ 
\
\section{Experimental Setup}

\begin{figure}
    \centering
    \includegraphics[width=1\linewidth]{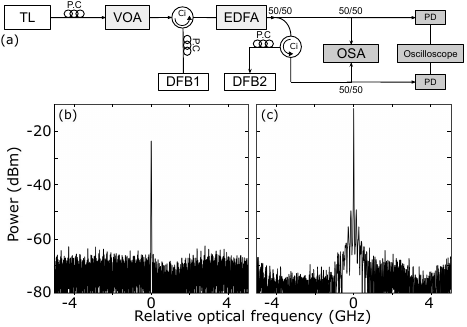}
    \caption{(a) Experimental setup for cascade injection: TL, tunable laser (Master); PC, polarization controller; VOA, variable optical attenuator; Ci, optical circulator; EDFA, erbium doped fiber amplifier; OSA, optical spectrum analyzer; DFB, laser diode; PD, photoreceiver. (b) Optical spectrum of the master laser and (c) the transmitter laser.}
    \label{setup}
\end{figure}

Figure \ref{setup}(a) presents a schematic of our experimental setup. The tunable laser (TL, Yenista TUNICS T100S) serves as the master laser, emitting an optical frequency of $\nu_M$. Its optical spectrum is shown in Fig. \ref{setup}(b).
 The signal then passes through a polarization controller (PC), and a variable optical attenuator (VOA) is employed to control the injection strength into the transmitter laser (DFB1), using a circulator (Ci). All the lasers used are directly fiber-coupled. DFB1 is a quantum well distributed feedback laser (LASERS-COM LDI-150-DFB-2.5G-20/70), emitting in a single mode at 1550 nm, with threshold current of 7.5 mA. Its optical spectrum is shown in Fig. \ref{setup}(c). Throughout the experiment, its optical frequency $\nu_T$ remains fixed.  
Chaos is generated by the optical injection of the TL into DFB1 and its properties are controlled by varying the injection strength, the frequency detuning defined as $\Delta \nu_{MT} = \nu_M - \nu_T$, and the pump current of the transmitter laser $I_T$. The signal of DFB1 is passed through an erbium-doped fiber amplifier (EDFA) to control the injection strength for the subsequent stages of the setup. The chaotic signal is injected into the receiver laser (DFB2) which is of the same type as DFB1.
Its frequency can be adjusted by varying the temperature, with a temperature coefficient of 
\begin{equation*}
\Delta \nu / \Delta T = -11.1 \, \text{GHz}
\end{equation*}
The detuning between DFB1 and DFB2 is varied   to measure the chaos synchronization in a range of frequencies larger than the chaos bandwidth. The temperature is adjusted in increments of $\Delta T \sim $ 0.1°, corresponding to a frequency shift of $\sim $ 1.1 GHz. \\ 
A 50/50 splitter enables us to analyze the transmitter laser chaos using an optical spectrum analyzer (Aragon-Photonics-BOSA400, with a spectral resolution of 10 MHz) and an oscilloscope (Lecroy LabMaster 36 Zi-A, sample rate 80 GS/s, 36 GHz bandwidth). The oscilloscope receives the output from two photodetectors (Newport 1474-A, 35 GHz bandwidth) that measure the dynamics of the DFB1 and DFB2 lasers. Since the measurement bandwidth of our setup (35 GHz) surpasses the bandwidth of the chaotic signals, we can perform a comprehensive analysis of the chaotic laser dynamics. Additionally, we measure the optical spectrum of the receiver laser and compare it to that of the transmitter laser.

\ 
\
\section{Results}

Our experimental setup allows us to control various parameters for the chaos generation: the frequency of the master laser $\nu_M$, the injection strength into the transmitter laser, and the pump current of the transmitter laser. Previous works have explored in detail the bifurcations to chaos in the optical injection system \cite{ohtsubo2017semiconductor,wiec_physicsreports}. We operate first in a negative detuning case where $\nu_M < \nu_T$. There the stationary frequency locked dynamics bifurcates to unlocked wave-mixing via a saddle node bifurcation and later to chaos by successive period doubling bifurcations. We operate with a low pump current in our transmitter laser ($I_T = 15$ mA) and a detuning $\Delta \nu_{MT} = -15 $ GHz. The receiver laser is pumped with 10 mA. The injection strength into the receiver laser is $\kappa_R = 0.16$, calculated as the ratio between the injection power into DFB2 and the emission power of DFB2 in free running operation.

\begin{figure}
    \centering
    \includegraphics[width=1\linewidth]{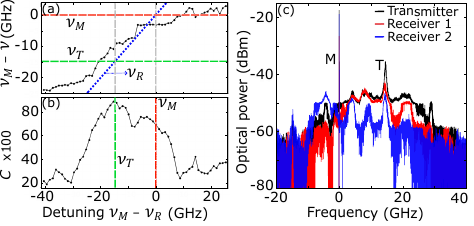}
    \caption{Results for negative detuning chaos  ($\Delta \nu_{MT} = -15$ GHz). (a) Nominal frequency (blue) and COM frequency (black) of receiver laser versus detuning. (b) Correlation between transmitter and receiver laser versus detuning. (c) Optical spectra of transmitter and receiver laser for $\Delta \nu_{MT} = -15$ GHz with master (M) and transmitter (T) laser frequencies. The spectrum of the receiver laser for $\Delta \nu_{MR} = -15$ GHz (-5 GHz) is shown in red (blue).}
    \label{chaos_negatif}
\end{figure}

We vary the temperature of DFB2 from 19°C to 25°C, corresponding to a change of the detuning $\nu_M - \nu_R$ from 25.07 GHz to -40.62 GHz. The change in $\nu_R$ under free-running conditions indicated in Fig. \ref{chaos_negatif}(a) by the dotted blue line.
The frequencies $\nu_M$ and $\nu_T$ are also indicated in Fig. \ref{chaos_negatif}(a) and (b) by the red and green dashed lines, respectively. The master laser frequency $\nu_M$ is the reference frequency, i.e., 0 GHz. 
We also calculate the evolution of the average optical frequency of the receiver laser under injection when varying the detuning $\nu_M - \nu_R$ [black line in Fig. \ref{chaos_negatif}(a)].
Globally, we can observe that this frequency evolves with $\nu_R$, since the chaos spectrum shifts together with the increasing detuning. This evolution becomes more significant when the frequency of the receiver laser $\nu_R$ enters the frequency range of the chaos spectrum (with $\nu_M - \nu_R \in \left[\ -30\ ;\ 7\ \right]$ GHz). 

The synchronization quality between the transmitter and the receiver lasers is evaluated by the correlation coefficient $C(\Delta t)$
\begin{equation}
    C(\Delta t) = \frac{<[P_T(t) - <P_T>][P_R(t+\Delta t) - <P_R>]>}{\sqrt{<[P_T(t) - <P_T>]^2><[P_R(t)-<P_R>]^2>}}
\end{equation}
where $P_T$ and $P_R$ are the output powers of the transmitter and receiver lasers, respectively. The time average is denoted by $< >$, and $\Delta t$ is a time shift. The value of $C$ is computed for each value of detuning between the master and the receiver lasers, and the result is presented in Fig. \ref{chaos_negatif}(b).

We find two parameter regions of high synchronization: the first, with the largest correlation value (greater than 89\% ), occurs when $\nu_M - \nu_R \simeq  -15$ GHz, or equivalently when $\nu_R \simeq  \nu_T$. The second region is found when $\nu_M - \nu_R \simeq 0$ GHz, or equivalently when $\nu_R \simeq \nu_M$. Figure \ref{chaos_negatif}(c) shows the optical spectra of the DFB1 and DFB2 lasers in the two synchronization regions. The peak labeled "M" stems from the master laser, and "T" stems from the transmitter laser. For $\nu_M - \nu_R = -15$ GHz the receiver laser optical spectrum has the best match with the transmitter spectrum, hence explaining the largest correlation coefficient achieved. As $\nu_R$ approches $\nu_M$, we find a second local maximum at  $\nu_M - \nu_R = -5$ GHz with somewhat lower correlation (76\%) than the maximum at -15 GHz.

\begin{figure}
    \centering
    \includegraphics[width=1 \linewidth]{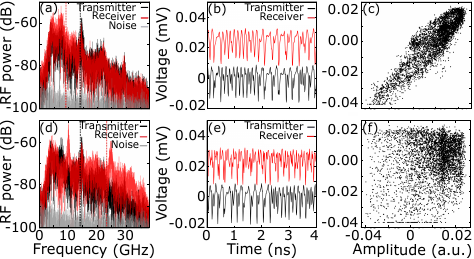}
    \caption{Temporal representation of the synchronization corresponding to the case in Figure 2, for two detuning case : (a),(b),(c) for $\Delta \nu_{MR} = -15.44$ GHz (90 \% correlation) and (d),(e),(f) for $\Delta \nu_{MR} = 9.74$ GHz (23 \% correlation). Panels (a),(d) show the RF power spectra for the noise floor (grey), transmitter chaos (black) and receiver chaos (red), with the respective bandwidth values indicated with the red and black dotted lines. Panels (b),(e) show the times series of the receiver chaos (red) with an offset of 0.02 mV, and the injected chaos (black), with the correlation plots (c) and (f).}
    \label{temporal}
\end{figure}

It is worth noting that, for all detunings shown in Fig. \ref{chaos_negatif}, the largest correlation coefficient is found for $\Delta t$ equal 45.53 ns, which corresponds to the optical length of the path from the transmitter laser to the receiver laser. Therefore synchronization between the transmitter and receiver laser occurs with zero lag. We have checked that the sliding-window cross-correlation \cite{slcc} shows no significant fluctuation of the correlation value along the time trace, meaning that no desynchronization events are detected.

\ 
\
We now study the synchronization in the time domain. In Fig. \ref{temporal}, we compare the example of best synchronization ($\nu_M - \nu_R = -15$ GHz) in Fig. \ref{temporal}(a-c) with an example of bad synchronization ($\nu_M - \nu_R = 9.74 $ GHz) in Fig. \ref{temporal}(d-f). RF spectra are shown in Fig. \ref{temporal}(a) and (d) for the transmitter laser (black), the receiver laser (red), and the noise floor (grey). The chaos bandwidth is the frequency range from DC up to the point where 80\% of the total power spectral density (PSD) of the chaotic signal is concentrated \cite{Bandwidth_1,Bandwidth_2}. It provides an indication of performance by capturing the spectral range where the majority of the signal power resides. In our case (same as Fig. \ref{chaos_negatif}), the values are 14.52 GHz for the transmitter laser chaos and 9.02 GHz for the receiver laser chaos, respectively. The best synchronization case, we find the same frequency signatures in the RF spectra of the transmitter and receiver lasers. This is not surprising in view of the good correspondence between the two optical spectra in Fig. \ref{chaos_negatif}(c). 
However, in the case of bad synchronization, the RF spectrum of the receiver laser is broader [Fig. \ref{temporal}(d)] and involves new frequencies initially absent in the transmitter laser spectrum due to the bandwidth enhancement induced by optical injection.  

In the case of best chaos synchronization, we can also note a good correspondence between the temporal waveforms, when a time-shift of 45.53 ns is applied corresponding to the optical path between the lasers. For $\nu_M - \nu_R = 9.74$ GHz, we indeed observe that receiver and transmitter temporal waveforms differ, with additional oscillations in the chaotic signal of the receiver laser, which explains the low correlation coefficient of 23\%.

\begin{figure}
    \centering
    \includegraphics[width=1\linewidth]{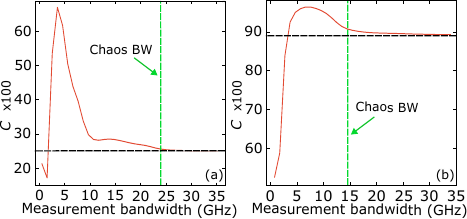}
    \caption{Effect on the MBW on the correlation coefficient for (a) bad synchronization ($\Delta \nu_{MT} = 9 $ GHz) and (b) best synchronization ($\Delta \nu_{MT} = -15.44 $ GHz). The green dashed lines indicate the CBW of the receiver laser chaos. The black dashed lines indicate the correlations for maximal MBW.}
    \label{bandwidth}
\end{figure}

Our experiment reveals a significant impact of the measurement bandwidth (MBW) on the measured chaos synchronization quality. As a reminder, in our experimental setup of Fig. \ref{setup}(a), the MBW is 35 GHz. In Fig. \ref{bandwidth} we demonstrate the effect on the correlation coefficient of numerically reducing the MBW by applying a low-pass filter on the measured time traces. We selected two cases of chaos synchronization, and calculated the correlation value by numerically modifying the measurement bandwidth. The first case examined in Fig. \ref{bandwidth}(a) represents a case of bad synchronization for a detuning $\nu_M - \nu_R = +9 $ GHz. In this case, the chaos bandwidth (CBW) of the receiver laser is 17.8 GHz. With a measurement bandwidth of 35 GHz, the correlation is 25\%. This value remains unchanged until the MBW approaches 25 GHz. When the MBW falls below this value, we observe an increase in the correlation, reaching a maximum of 66\%. The second case explores the case of the best synchronization with a correlation coefficient of 90\% (for $\nu_M - \nu_R = -15.44$ GHz) and a CBW of about 9.02 GHz. Similarly by reducing the MBW, we notice an increase in the correlation value up to 96\%. We conclude that a MBW well above the CBW is necessary to avoid an overestimation of the chaos synchronization quality.
\ 
\

\begin{figure}
    \centering
    \includegraphics[width=1 \linewidth]{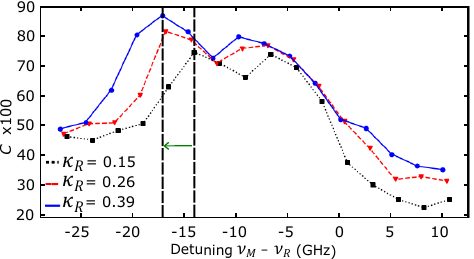}
    \caption{Correlation coefficient of chaos synchronization as function of detuning for injection strengths $\kappa_R$ = 0.15 (black dotted line), 0.26 (red dashed line), and 0.39 (blue solid line), for $\Delta \nu_{MT} = -15$ GHz. The two vertical black lines indicate the shift of the correlation peaks with increasing $\kappa_R$.}
    \label{parametric}
\end{figure}

Figure \ref{parametric} further analyzes the impact on chaos synchronization of the injection strength from the transmitter laser to the receiver laser. We control the injection strength with the EDFA, and the injection strength is fixed during the chaos synchronization measurement. We study the same case as shown in Fig. \ref{chaos_negatif} with negative detuning between the master and the transmitter laser ($\Delta \nu_{MT} = -15$ GHz ), and we make three measurements with fixed parameters ($I_T = 15$ mA and $I_R = 15$ mA), for three different injection strengths and an emission power of 1.7 mW for the receiver laser. We draw several conclusions. Firstly, the evolution of the correlation coefficient versus $\nu_M - \nu_R$ remains qualitatively the same for all three cases, with two parameters regions of good synchronization. Secondly, as the injection strength increases, the maximum correlation coefficient also increases, reaching up to 90 \%. We can see that for the case of 0.25 mW injection power ($\kappa_R = 0.15$) represented by the black line, the maximum correlation is lower than the 0.45 mW case ($\kappa_R = 0.26$) denoted by the blue line. Thirdly, we note that the maximum  correlation coefficient occurs for a smaller $\nu_M - \nu_R$ with increasing injection strength. As a matter of fact, an increase in the injection strength leads to a reduction of carrier density, which also has an impact on the refractive index $n_g$ and, consequently, on the resonance frequency of the receiver laser \cite{frequencyshift}, according to  $\Delta \nu_R = - \frac{\nu_R}{n_g} \frac{dn_e}{dN} \Delta N$, where $\Delta \nu_R$ is the resonance frequency shift of the receiver, $n_g$ is the group refractive index, and $dn_e / dN$ the incremental change in the effective index, caused by the carrier density modification $\Delta N$.
Therefore, an increase in injected power induces an increase in the refractive index and thus a decrease in $\nu_R$. Since the maximum correlation coefficient for chaos synchronization is reached for $\nu_M \simeq \nu_R$ or $\nu_T \simeq \nu_R$, and both $\nu_M$ and $\nu_T$ remain fixed, decrease of the actual $\nu_R$ induced by the higher injection strength must be compensated by an increase of the nominal $\nu_R$, and thus the best synchronization is found for smaller detunings $\nu_M - \nu_R$ as $\kappa_R$ increases. This is the same mechanism that causes the injection locking range to shift to smaller detunings as the injection strength increases.

\begin{figure}
    \centering
    \includegraphics[width=1 \linewidth]{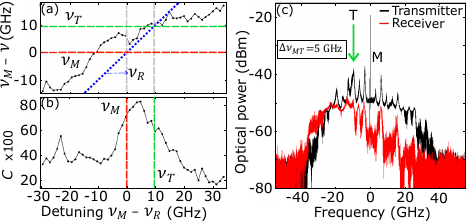}
    \caption{ In the case of positive detuning chaos ($\Delta \nu_{MT} = 9.5$ GHz ) : (a) Evolution of the receiver chaos frequency and (b) the correlation between the two chaotic signals as function of the detuning between master and receiver laser. (c) shows the optical spectra of the two chaos for $\nu_M - \nu_R = 5$ GHz, indicating the master laser (M) and transmitter (T) laser frequencies.}
    \label{chaos_positif}
\end{figure}

\ 
\
Finally, we study the case of chaos generated by positive detuning between master and transmitter lasers. To reach this chaotic regime, it is necessary to change our laser parameters, keeping the same injection strength as in the case of Fig. \ref{chaos_negatif}. We fixe the pump current of the transmitter laser to a higher value $I_T = 30$ mA, and we shift $\nu_M$ to achieve a detuning value of $\Delta \nu_{MT} = 9.5 $ GHz. For these parameters, it is possible to generate a new type of chaos. By adjusting the detuning between the master and transmitter lasers, the initial stationary regime of the transmitter laser evolves towards an oscillating state, passing through a Hopf bifurcation point and secondary bifurcations when increasing the injection strength, finally leading to chaos. Here, the CBW of the transmitter laser is 11 GHz.  As previously, we plot the evolution of the frequency of the receiver laser [Fig. \ref{chaos_positif}(a)] and the correlation coefficient between the transmitter laser and the receiver laser chaotic signals versus $\nu_M - \nu_R$ [Fig. \ref{chaos_positif}(b)]. We observe a single region with high correlation for $\nu_M - \nu_R$ between 0 and 10 GHz or equivalently when $\nu_R$ is between $\nu_R \simeq \nu_M$ and $\nu_R \simeq \nu_T$. This single correlation region for chaos generated by positive $\Delta \nu_{MT}$ contrasts with the two high correlation regions found for chaos generated by negative $\Delta \nu_{MT}$ [Fig. \ref{chaos_negatif}]. When comparing with Fig. \ref{chaos_negatif}, we note a weaker correspondence in the optical spectra of the master and receiver lasers [Fig. \ref{chaos_positif}(c)] at the detuning of maximal correlation. The optical spectra of the transmitter and receiver lasers differ, although the mean chaos frequency in the receiver laser is close to both $\nu_T$ and $\nu_M$.
\section{Conclusion}
In conclusion, in contrast to the classical two laser synchronization scheme, where the best synchronization is achieved when the two laser frequencies are almost equal, the addition of a third laser allows to identify three scenarios of good synchronization, that is when either $\nu_R \simeq \nu_M$,  $\nu_R \simeq \nu_T$ for $\Delta \nu_{MT} < 0$ or $\nu_R$ in between $\nu_M$ and $\nu_T$ for $\Delta \nu_{MT} > 0$. Which scenario of synchronization is taking place can be adjusted by varying both the detuning between the receiver and master lasers and the injection strength of the receiver laser. The presented experimental setup enables the generation of different chaos patterns stemming from various bifurcations, each with different bandwidths. Across all cases, we achieved a 90\% correlation, with the profile depending on the parameters of the three lasers. \\
The possibility to adjust the synchronization quality when varying the system parameters is essential to ensure the security in a cryptography application. Here the possibility to switch from the synchronization property seen in Fig. \ref{chaos_positif} to the one of Fig. \ref{chaos_negatif} is a clear advantage of this setup using three cascaded lasers and has no counterpart in the classical master - receiver chaos synchronization scheme.
We also emphasize the importance of the measurement bandwidth on correlation values and caution against the interpretation of the synchronization quality if the measurement bandwidth is smaller than the chaos bandwidth. Compared to some other papers, our higher MBW and CBW (due to optical injection) allows us to achieve different results, thus offering a fresh perspective on chaos synchronization.


\textbf{Funding} The Chair in Photonics is supported by Region Grand Est, GDI Simulation, Departement de la Moselle, European Regional Development Fund, CentraleSupelec, Fondation CentraleSupelec, and Eurometropole de Metz. \\ 

\textbf{Data Availability Statement} Data underlying the results presented in this paper are
not publicly available at this time but may be obtained from the authors upon
reasonable request. \\ 

\textbf{Disclosures} The authors declare no conflicts of interest.

\bibliography{apssamp}
\end{document}